\documentclass[epj]{svjour}
\usepackage{graphics}
\begin{document}
\title{Scattering of Stark-decelerated OH radicals with rare-gas atoms}
\author{Ludwig Scharfenberg\inst{1} \and Koos B.
Gubbels\inst{1} \inst{2} \and Moritz Kirste\inst{1} \and Gerrit C.
Groenenboom\inst{2} \and Ad van der Avoird \inst{2} \and Gerard
Meijer\inst{1} \and Sebastiaan Y.T. van de Meerakker\inst{1}
}                     
%
%
\institute{Fritz-Haber-Institut der Max-Planck-Gesellschaft,
Faradayweg 4-6, 14195 Berlin, Germany \and Institute for Molecules
and Materials, Radboud Universiteit Nijmegen, Heyendaalseweg 135,
6525 AJ Nijmegen, The Netherlands}
\date{Received: date / Revised version: date}
%

\abstract{ We present a combined experimental and theoretical study
on the rotationally inelastic scattering of OH ($X\,^2\Pi_{3/2},
J=3/2, f$) radicals with the collision partners He, Ne, Ar, Kr, Xe,
and D$_2$ as a function of the collision energy between $\sim 70$
cm$^{-1}$ and 400~cm$^{-1}$. The OH radicals are state selected and
velocity tuned prior to the collision using a Stark decelerator, and
field-free parity-resolved state-to-state inelastic relative
scattering cross sections are measured in a crossed molecular beam
configuration. For all OH-rare gas atom systems excellent agreement
is obtained with the cross sections predicted by close-coupling
scattering calculations based on accurate \emph{ab initio} potential
energy surfaces. This series of experiments complements recent
studies on the scattering of OH radicals with Xe [Gilijamse \emph{et
al.}, Science {\bf 313}, 1617 (2006)], Ar [Scharfenberg \emph{et
al.}, Phys. Chem. Chem. Phys. {\bf 12}, 10660 (2010)], He, and D$_2$
[Kirste \emph{et al.}, Phys. Rev. A {\bf 82}, 042717 (2010)]. A
comparison of the relative scattering cross sections for this set of
collision partners reveals interesting trends in the scattering
behavior.
\PACS{
      {PACS-key}{discribing text of that key}   \and
      {PACS-key}{discribing text of that key}
     } 
} 

\maketitle

\section{Introduction}
The study of collisional energy transfer between simple atoms and
molecules has been essential for our present understanding of the
dynamics of molecular interactions, and for testing our ability to
accurately calculate potential energy surfaces that govern these
interactions \cite{Levine:reaction-dynamics}. Rotationally inelastic
scattering is one of the simplest scattering processes, and has been
studied with ever increasing level of detail during the last
decades. Experimental studies at a full state-to-state level are
nowadays possible, revealing detailed information on the potential
energy surfaces and the resulting motion on these surfaces
\cite{Schiffman:IRPC14:1995,Liu:IRPC9:187}.

Rotationally inelastic scattering of free radical species such as OH
or NO with atomic collision partners has been of special interest in
molecular scattering experiments
\cite{Chandler:book,Whitehead:RepProgPhys59:993}. The scattering of
these open-shell species in a $^2\Pi$ electronic state involves more
than one Born-Oppenheimer potential surface, resulting in rich
multi-surface dynamics with various quantum interference effects
\cite{Kohguchi:ARPC98:421}. At a full state-to-state level,
collision induced transitions between rotational, spin-orbit, and
$\Lambda$-doublet levels have been studied
\cite{Schreel:JCP99:8713,Beek:JCP113:628,Leuken:JPC99:15573}.
Sophisticated beam production and product state detection methods
have been developed to measure differential cross sections
\cite{Kohguchi:Science294:832,Gijsbertsen:JCP123:224305}, the steric
asymmetry of the collision \cite{Stolte:NAT353:391}, and the
alignment or orientation of the collision products
\cite{Beek:PRL86:4001,Paterson:JCP129:074304,Paterson:PCCP11:8804}.
The wealth of scattering data that is available for these systems,
together with the spectroscopic data of the bound states of relevant
complexes \cite{Heaven:ARPC43:283,Heaven:IRPC24:375}, offers
stringent tests for \emph{ab initio} potential energy surfaces
(PES's) and for quantum scattering calculations.

In recent years, new approaches to perform high-precision inelastic
scattering experiments involving radical species have become
possible with the development of the Stark-deceleration technique
\cite{Meerakker:NatPhys4:595}. The Stark deceleration method
exploits the concepts of charged-particle accelerator physics to
produce molecular beams with a tunable velocity and almost perfect
state purity \cite{Bethlem:PRL83:1558}. The method was first applied
to molecular scattering studies in 2006, when a Stark-decelerated
beam of OH ($X\,^2\Pi_{3/2}, J=3/2, f$) radicals was scattered with
a conventional beam of Xe atoms \cite{Gilijamse:Science313:1617}. By
tuning the velocity of the OH radicals between 33~m/s and 700~m/s
prior to the collision, the center-of-mass collision energy was
varied between 60 cm$^{-1}$ and 400 cm$^{-1}$. This energy range
encompasses the energetic thresholds for inelastic scattering to the
first excited rotational levels of the OH radical, and the threshold
behavior of the inelastic state-to-state cross sections was
accurately determined. Excellent agreement was found with cross
sections derived from coupled channel calculations on \emph{ab
initio} computed potential energy surfaces.

Since this first proof-of-principle experiment, a new crossed beam
scattering apparatus was developed that employs an improved version
of the Stark decelerator. With this decelerator, packets of OH
radicals can be produced with a superior number density, a narrower
velocity spread, and a higher quantum state purity
\cite{Scharfenberg:PRA79:023410}. This apparatus enables
state-to-state scattering experiments as a function of the collision
energy with a sensitivity that exceeds that of conventional crossed
beam scattering experiments. This was demonstrated first on the
benchmark OH ($X\,^2\Pi$)-Ar system, for which parity-resolved
integral state-to-state scattering cross sections for in total
13~inelastic scattering channels have been determined as a function
of the collision energy \cite{Scharfenberg:PCCP12:10660}. Recently,
the same methodology was applied to the scattering of OH radicals
with He atoms and D$_2$ molecules \cite{Kirste:PRA82:042717}.

These experiments challenge the most accurate potential energy
surfaces and quantum scattering calculations presently available.
For the scattering of OH radicals with Ar and He atoms, excellent
agreement was found between experiment and theory, although at high
collision energies and for specific inelastic channels deviations
were found. For the OH-He system, the almost perfect quantum state
purity offered by the Stark decelerator enabled the observation of
the strong propensities for preferred excitation into final states
of certain parity that had been predicted for this system
\cite{Schreel:JCP99:8713}.

Here we report new measurements on the rotationally inelastic
scattering of OH ($X\,^2\Pi_{3/2}, J=3/2, f$) radicals with Ne, Kr,
and Xe atoms at collision energies between 60~cm$^{-1}$ and
400~cm$^{-1}$. The measured cross sections for the OH-Xe system
confirm the cross sections that were determined for this system in
our earlier work \cite{Gilijamse:Science313:1617}. For all three
systems, excellent agreement is obtained with cross sections that
are derived from quantum scattering calculations based on available
potential energy surfaces.

These studies complement our previous investigations on the
scattering of OH radicals with He atoms, Ar atoms and D$_2$
molecules, and together form a complete data set on the scattering
of OH radicals with rare-gas atoms. A comparison of the relative
scattering cross sections for the various collision partners is
presented that reveals interesting trends in the scattering
behavior.

\section{Experiment}\label{sec:experiment}

\subsection{Experimental setup}\label{sec:2:1}
\begin{figure}
    \centering
 \includegraphics{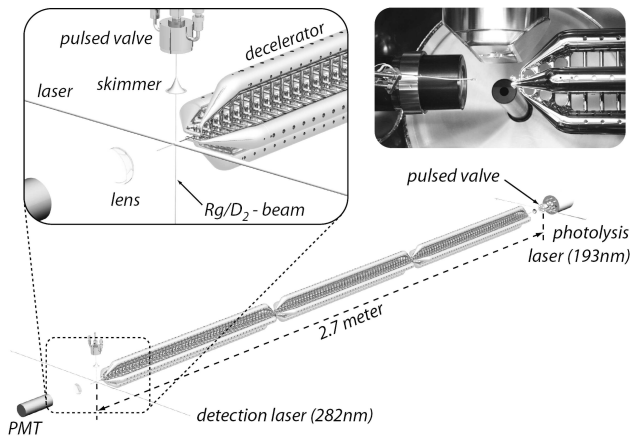}
    \caption{
                        Scheme of the experimental setup.
                        A pulsed beam of OH radicals is produced by laser photolysis of HNO$_3$ seeded in
                        either Kr or Ar. The radicals pass a 2.6 m long Stark decelerator and are scattered by a
                        pulsed beam of rare-gas (Rg) atoms or D$_2$ molecules. The OH radicals are state-selectively
                        detected using a laser-induced fluorescence scheme. The fluorescence is imaged onto a
                        photo multiplier tube by a lens. The upper right inset shows a photograph of the
                        beam crossing region with lens holder, light baffle, and the exit of the Stark decelerator. The secondary beam
                        source is protruding from the top.
                    }
    \label{fig:setup}
\end{figure}

The experiments are performed in a crossed molecular beam apparatus
that is schematically shown in Figure \ref{fig:setup}. A detailed
description of this machine, as well as of the production, Stark
deceleration, and detection of OH radicals can be found in Ref.
\cite{Scharfenberg:PRA79:023410,Scharfenberg:PCCP12:10660}; only the
most essential aspects of the experiment are described here.

A pulsed supersonic beam of OH radicals is created by photolysis of
nitric acid seeded in either krypton or argon. After the supersonic
expansion nearly all OH radicals reside in the lowest rotational ($J
= 3/2$) and vibrational level of the $X ^2\Pi_{3/2}$ electronic
ground state. The two $\Lambda$-doublet components of this level,
labeled $e$ and $f$, are populated equally in the beam since their
energy difference is only 0.05~cm$^{-1}$. OH radicals that reside in
the energetically higher lying $f$-component can be focused and
velocity tuned using the Stark decelerator, whereas OH radicals in
the $e$-component are deflected from the beam axis. After passing a
skimmer, the packet of OH radicals enters the $3 \times 3$ mm$^2$
opening of the decelerator. A sequence of high voltage pulses is
applied to the decelerator electrodes to generate time-dependent
electric field configurations that either decelerate, guide or
accelerate the OH radicals. The Stark decelerator that is used here
employs the so-called $s=3$ operation mode
\cite{Meerakker:PRA71:053409,Meerakker:PRA73:023401}, and has been
described in detail in Ref. \cite{Scharfenberg:PRA79:023410}.

The packet of OH ($X ^2\Pi_{3/2}, J=3/2, f$) radicals that emerges
from the decelerator has a quantum state purity of $> 99$\% and
intersects the central axis of the secondary beam under 90$^\circ$
at a distance of 16.5~mm from the decelerator exit. Collisions take
place in a field free region and the initially uneven distribution
over $M_J$-components which is present inside the decelerator is
assumed to be scrambled completely while the molecules move towards
the collision region.

A temperature controlled solenoid valve produces the secondary beam
of rare-gas atoms or D$_2$ molecules. The mean forward velocity of
this beam is inferred from a time-of-flight measurement using two
microphone-based beam detectors that are placed 300~mm apart. To
ensure single-collision conditions, the intensity of the secondary
beam is kept sufficiently low so that the decrease of the initial
population in the $J=3/2,f$ level remains below 4 percent.

The collision products are state-selectively detected \emph{via}
saturated laser-induced fluorescence when the most intense part of
the OH packet is in the center of the collision region. A pulsed dye
laser is used to induce various rotational transitions of the $A
^{2}\Sigma^+, v=1 \leftarrow X ^2\Pi, v=0$ band. The laser beam
intersects both molecular beams under 90$^\circ$ and the
off-resonant fluorescence is collected by a lens and imaged onto a
photomultiplier tube (see the inset of Figure \ref{fig:setup}).

\subsection{Measurement procedure and data analysis}\label{sec:2:2}
The collision energy range between 60~and 400~cm$^{-1}$ is covered
using two measurement intervals with overlapping energy, as
described in Ref. \cite{Scharfenberg:PCCP12:10660}. For these
intervals, molecular beams of OH radicals are produced using Kr and
Ar as seed gases, resulting in OH radical beams with mean initial
velocities of 430~m/s and 615~m/s, respectively. Within each
interval, the collision energy is varied by tuning the velocity of
the OH radicals prior to the collision using the Stark decelerator.
For collisions with Ne, Ar, Kr and Xe atoms, the mean velocity of
the rare-gas atom beam is kept constant for all measurements. The
mean speed and corresponding valve temperatures for the different
rare-gas beams are: 445~m/s (Ne, 93~K), 400~m/s (Ar, 110~K), 330~m/s
(Kr, 163~K) and 300~m/s (Xe, 220~K). For collisions with He atoms
and D$_2$ molecules, the small reduced mass makes it inconvenient to
vary the collision energy by solely tuning the speed of the OH
packet. Therefore, the speed of the secondary beam is varied in
comparatively large steps by changing the temperature of the valve.
For a given valve temperature, the OH velocity is then tuned from
168 to 741~m/s. The D$_2$ molecules are assumed to be distributed
between ortho and para rotational levels according to the
statistical weights, i.e., 67\% of the molecules are expected to
populate rotational levels with $J$ even and 33\% levels with $J$
odd.

The Stark decelerator provides packets of OH radicals at a rate of
10~Hz. The secondary beam is operated at 5~Hz and collision signals
are inferred from the fluorescence intensity difference between
alternating shots of the experiment. The collision energy dependence
of the scattering channels is measured \emph{via} a quasi-continuous
cycle, as explained in Ref. \cite{Scharfenberg:PCCP12:10660}. The
collision signals are obtained from typically 1000 runs of the
experiment, and quoted error bars represent the statistical
fluctuation of the measured mean values. Both photon counting and
analog detection are used in the data acquisition
\cite{Scharfenberg:PCCP12:10660}.

Our experiment is not sensitive to elastic scattering; only
scattering events that change the internal quantum state of the OH
radical can be detected. Within the studied collision energy range,
collisional excitation to at most 13 rotational levels can occur.
These levels are labeled as $F_i(Je/f)$, where $i = 1$ denotes the
$X\,^2\Pi_{3/2}$ and $i=2$ the $X\,^2\Pi_{1/2}$ spin-orbit
manifolds, and the parity labels $e$ and $f$ correspond to the two
$\Lambda$-doublet components of each rotational level. An energy
level diagram with all relevant rotational levels is shown in Figure
\ref{fig:F1_channels}. Note that the $\Lambda$-doublet splitting is
largely exaggerated in this figure for reasons of clarity. The
rotational transitions that are used to probe the individual levels,
as well as the excitation rates that are used to convert measured
fluorescence intensities to populations, are specified in
\cite{Scharfenberg:PCCP12:10660}.

The experimental scattering signals are most easily compared with
theoretical calculations when relative inelastic scattering cross
sections are derived from the observations. The relative scattering
cross section for a specific channel is proportional to the total
number of molecules that is detected in the corresponding quantum
state. However, the detection volume is necessarily limited and in
general not all molecules can be detected. A density-to-flux
transformation is required to relate the measured relative
populations in final states to relative scattering cross sections.
Under our experimental conditions, the resulting correction is small
\cite{Scharfenberg:PCCP12:10660}. For the scattering of OH with Ne,
Ar, Kr and Xe atoms, we have performed the transformation using the
differential cross sections determined from theory. For He and
D$_2$, the density-to-flux correction can be safely omitted due to
the small mass of the collision partner compared to the mass of the
OH radical.

\section{Theory}\label{sec:theory}

The theory for the scattering of ${}^2 \Pi$-state molecules with
${}^1 S$-state atoms is well established
\cite{Alexander:JCP76:5974}. In particular, collision studies
between OH molecules and rare-gas atoms at low collision energies
have received a lot of attention over the last years
\cite{Gilijamse:Science313:1617,Lara:PRL97:183201,Sanchez:PRA73:022703,Pavlovic:JPCA113:14670}.
In this section we only give a brief summary of the relevant theory.
A more extensive account can be found in Ref.
\cite{Tscherbul:FD142:127}.

The Hamiltonian that describes the scattering of ground state OH
($X^2\Pi$) with rare-gas atoms is given by
\begin{equation}\label{eqham}
\hat{H}=\frac{-\hbar^2}{2 \mu R}\frac{\partial^2}{\partial
R^2}R+\frac{\hat{L}^2}{2\mu
R^2}+\sum_{\Lambda',\Lambda}|\Lambda'\rangle{V}_{\Lambda',\Lambda}(R,\theta)\langle\Lambda|+\hat{H}_{\rm
OH},
\end{equation}
where $R$ is the length of the vector ${\bf R}$ that connects the
center-of-mass of the OH molecule and the rare-gas atom, $\mu$ is
the reduced mass of the atom-OH complex, $\hat{L}$ is the angular
momentum operator corresponding to end-over-end rotation of the
OH-rare gas atom complex, and $\hat{H}_{\rm OH}$ is the Hamiltonian
of the OH molecule in the ($X^2\Pi$) ground state. The $X\,^2\Pi$
electronic ground state of the OH radical has two degenerate
components with projections $\Lambda=\pm 1$ of the orbital
electronic angular momentum on the internuclear $\hat{r}$-axis. The
OH-rare gas interaction is represented by the operators
$|\Lambda'\rangle{V}_{\Lambda',\Lambda}(R,\theta)\langle\Lambda|$
that couple different electronic states $\Lambda$ and $\Lambda'$.
The angle $\theta$ defines the angle between the unit vector
$\hat{R}$ and the OH bond direction $\hat{r}$, with $\theta=0$
corresponding to collinear atom-HO. The Hamiltonian of OH includes
rotation, spin-orbit coupling and $\Lambda$-doubling, where we use
the OH rotational constant $B = 18.5487$ cm${}^{-1}$, the spin-orbit
coupling constant $A = -139.21$ cm${}^{-1}$, and $\Lambda$-doubling
parameters $p = 0.235$ cm${}^{-1}$ and $q = -0.0391$ cm${}^{-1}$
\cite{Maillard:JMolSpec63:120}. From Eq. (\ref{eqham}) it follows
that differences in the collisions between OH and the various rare
gas atoms originate from the differences in the interaction
potential and the reduced mass.

When the rare-gas atom approaches the OH molecule, the electronic
degeneracy of the $\Pi$ state is lifted. The resulting matrix
elements $V_{\Lambda,\Lambda'}$ of the potential  are nonzero for
$\Lambda' - \Lambda=0, \pm 2$, and two potential energy surfaces are
involved in the scattering process. The potential energy surfaces
can be expanded in Racah normalized spherical harmonics
\begin{eqnarray}\label{eqpot}
V_{1,1}  = V_{-1,-1} &=& \frac{V_{A'}+V_{A''}}{2} = \sum_l v_{l,0}(R)C_{l,0}(\theta,0),\nonumber \\
V_{1,-1} = V_{-1,1}  &=& \frac{V_{A''}-V_{A'}}{2} = \sum_l
v_{l,2}(R)C_{l,2}(\theta,0),
\end{eqnarray}
where $A'$ and $A''$ refer to the reflection symmetry of the
electronic states. The surfaces $V_{1,1}$ and $V_{1,-1}$ are often
referred to as the sum $V_{\rm{sum}}$ and difference $V_{\rm{diff}}$
potential energy surface, respectively.

\emph{Ab initio} calculations for the OH-atom interaction energy can
be performed using the MOLPRO program package \cite{Molpro:2008},
which has resulted in potentials for the OH-He
\cite{Lee:JCP113:5736}, OH-Ne
\cite{Lee:JCP113:5736,Sumiyoshi:PCCP12:8340}, OH-Ar
\cite{Scharfenberg:PCCP12:10660}, OH-Kr
\cite{Sumiyoshi:MolPhys108:2207} and OH-Xe
\cite{Gilijamse:Science313:1617} complexes. The most relevant
properties of the various interacting systems, such as the reduced
mass of the OH-rare gas atom complex, the polarizability of the rare
gas atom, the minima of the potential energy surfaces in the two
different linear configurations of the complex ($\theta = 0^{\circ}$
and $\theta=180^{\circ}$), as well as the position and energy of the
minimum of the $A'$ potential at a nonlinear geometry, are listed in
Table \ref{tabpot}.

The potential energy surfaces vary in the quality of the basis set
used, and the quality of the method. All potentials used the
counterpoise procedure to correct for the basis set superposition
error \cite{Boys:MolPhys19:553}. The OH-He and the OH-Ne potentials
of Lee {\it et al.} were both calculated with a spin-restricted
coupled-cluster method with single and double excitations and
perturbative triples [RCCSD(T)]. The augmented triple-zeta
correlation-consistent basis set (aug-cc-pVTZ) was used with an
additional (3$s$, 3$p$, 2$d$, 2$f$, 1$g$) set of bond functions
centered in the midpoint of the vector ${\bf R}$
\cite{Lee:JCP113:5736}. The Ne-OH potential was also calculated more
recently by Sumiyoshi {\it et al.}, who used an explicitly
correlated, spin-unrestricted approach [UCCSD(T)-F12b] with a larger
quintuple-zeta basis set (aug-cc-pV5Z) \cite{Sumiyoshi:PCCP12:8340}.
Although we calculated the cross sections for both Ne-OH potentials,
we only show the results with the most recent potential, which gave
a clearly better agreement with experiment. This is probably due to
the larger basis set and the improved calculation method, which
includes explicit electron correlations that particularly enhance
the accuracy of the short-ranged behavior of the potential. For the
Xe-OH potential, RCCSD(T) was used with a qaudruple-zeta basis set
(aug-cc-pVQZ) and with a set of (3$s$, 3$p$, 2$d$, 1$f$, 1$g$)
mid-bond orbitals with geometry-dependent exponents
\cite{Gilijamse:Science313:1617}. For the Kr-OH potential, the
UCCSD(T)-F12b approach was used with the aug-cc-pVQZ basis set
\cite{Sumiyoshi:MolPhys108:2207}. Finally, the Ar-OH potential
surface was calculated with a spin-unrestricted approach [UCCSD(T)],
where the basis set was extrapolated to the complete basis set
limit, and where also an averaging over the $v=0$ motion of the OH
molecule was performed \cite{Scharfenberg:PCCP12:10660}. For the
other systems, the OH molecular geometry was assumed frozen at its
equilibrium bond length (OH-He, Ne, Kr) or at its vibrationally
averaged distance (OH-Xe).

In order to calculate the OH monomer eigenfunctions it is convenient
to use  a parity adapted Hund's case (a) basis set, labeled by
$|\Omega,J,M_J,p \rangle$ with $J$ the total angular momentum of the
OH molecule, $\Omega$ and $M_J$ the projections on the molecular and
space-fixed quantization axes, and $p$ the parity under inversion.
For the exact OH eigenfunctions, $|\Omega|$ is nearly a good quantum
number. The total angular momentum of the OH-atom complex is
represented by the operator
$\hat{\mathbf{F}}=\hat{\mathbf{J}}+\hat{\mathbf{L}}$, whose
eigenfunctions are obtained by coupling the monomer basis with the
spherical harmonics $|L,M_L\rangle=Y_{L,M_L}(\vartheta,\varphi)$,
where $\vartheta$ and $\varphi$ are the space-fixed spherical
coordinates of the vector ${\bf R}$. Assuming that the OH bond
length is fixed, we write the scattering wave functions as products
of radial and angular functions,
\begin{equation} \label{eqbas}
\Psi^{F,M_F,\mathcal{P}}_{\beta,L} =\frac{1}{R} \sum_{\beta',L'}
\chi^{F,M_F,\mathcal{P}}_{\beta',L'\leftarrow
\beta,L}(R)\psi^{F,M_F,\mathcal{P}}_{\beta',L'}(\hat{R},\hat{r}),
\end{equation}
where $\beta$ is a shorthand notation for the monomer quantum
numbers $(F_i,J)$ with $i$ to distinguish between the $F_1$ and
$F_2$ spin-orbit manifolds of the OH eigenstates. Note that the
total angular momentum $F$, its space-fixed projection $M_F$ and the
parity of the complex $\mathcal{P} = p (-1)^L$ are conserved in the
collision process. The experimentally relevant scattering
properties, i.e. the cross sections, are conveniently expressed in
terms of the scattering matrix, which can be obtained using standard
asymptotic matching procedures \cite{Johnson:JCompPhys13:445}. The
obtained $S$-matrix is then related to the scattering amplitudes,
which in turn determine the differential cross sections
\cite{Child:Book}.

In order to achieve convergence of the calculated cross-sections, we
used a basis set that included all OH rotational states up to an
angular momentum of $J=21/2$, and took into account all partial wave
contributions up to a total angular momentum of $F=241/2$. For the
propagation of the wavefunction, the renormalized Numerov method was
used, starting at $4 a_0$ and continuing to $35 a_0$ with  $a_0$ the
Bohr radius. The cross sections were evaluated on an energy grid
with a 5 cm${}^{-1}$ interval spacing, well below the experimental
energy resolution in all cases. It is noted that this energy grid is
too sparse, however, to identify individual scattering resonances
that occur at collision energies around the energetic thresholds.

\begin{figure}
    \centering
        \includegraphics{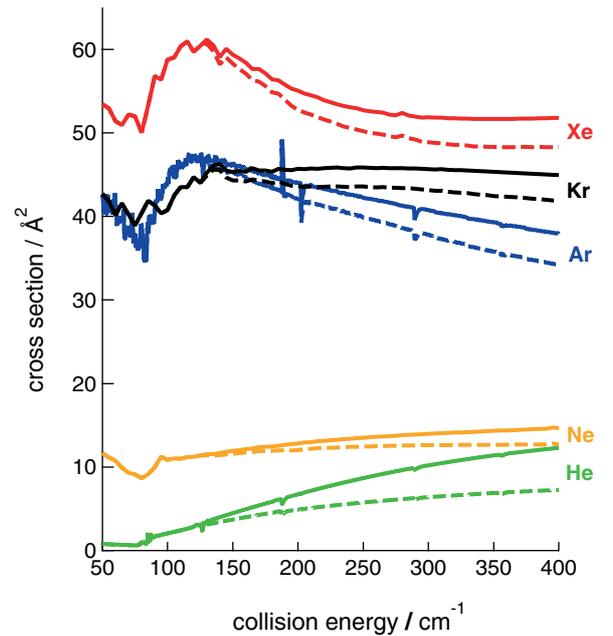}
    \caption{The calculated total integral cross sections for inelastic scattering of OH ($X\,^2\Pi_{3/2}, J=3/2, f$) radicals with He, Ne, Ar, Kr and Xe atoms (solid lines).
    The contribution
    of transitions into the $F_1$ spin-orbit manifold are shown by the dashed lines.}
    \label{fig:total_inelastic}
\end{figure}

In Fig. \ref{fig:total_inelastic} the total integral inelastic cross
sections (the sum of the integral cross sections over all inelastic
channels) are shown for collisions of OH ($X\,^2\Pi_{3/2}, J=3/2,
f$) radicals with the five different rare-gas atoms. In this figure,
the contribution to the total cross section of collisions that
populate levels within the $F_1$ manifold are indicated. The total
inelastic cross section is seen to rise with increasing reduced
mass, increasing atom polarizability and increasing well depth of
the potential. The total inelastic cross sections as presented in
Figure \ref{fig:total_inelastic} can be used to deduce absolute
state-to-state inelastic cross sections from the experimentally
determined relative state-to-state cross sections. These are
presented in the next section.

\begin{table}\label{tabpot}
            \renewcommand{\arraystretch}{1.5}
            \renewcommand{\tabcolsep}{0.1cm}
            \begin{tabular}{@{}l c c c c c c c c c c c l@{}}
                \hline
                atom & \verb+ + & $\mu$ & \verb+ + & $\alpha$ & \verb+ + & $E$& \verb+ + & $R$ & \verb+ + & $\theta$ & \verb+ + & PES\\
                \hline\hline
                \textbf{He} & & 3.24      & &  0.21  & & -27.1  & & 6.54 & & 0    & & $A^{\prime}$, $A^{\prime\prime}$\\
                                        & &           & &        & & -21.8  & & 6.09 & & 180  & & $A^{\prime}$, $A^{\prime\prime}$\\
                                        & &           & &        & & -30.0  & & 5.69 & & 68.6 & & $A^{\prime}$         \\
                \textbf{Ne} & & 9.22      & &  0.40  & & -59.34 & & 6.53 & & 0    & & $A^{\prime}$, $A^{\prime\prime}$\\
                                        & &           & &        & & -45.18 & & 6.14 & & 180  & & $A^{\prime}$, $A^{\prime\prime}$\\
                                        & &           & &        & & -59.60 & & 5.82 & & 67.6 & & $A^{\prime}$     \\
                \textbf{Ar} & & 11.9      & &  1.64  & & -141.7 & & 7.01 & & 0    & & $A^{\prime}$, $A^{\prime\prime}$\\
                                        & &           & &        & & -92.4  & & 6.70 & & 180  & & $A^{\prime}$, $A^{\prime\prime}$\\
                                        & &           & &        & & -137.1 & & 6.18 & & 74.8 & & $A^{\prime}$     \\
                \textbf{Kr} & & 14.14       & &  2.48    & & -172.5 & & 7.2  & & 0    & & $A^{\prime}$, $A^{\prime\prime}$\\
                                        & &           & &        & & -110.3 & & 6.8  & & 180  & & $A^{\prime}$, $A^{\prime\prime}$\\
                                        & &           & &        & &  -177.0 & &  6.25 & &  78 & & $A^{\prime}$     \\
                \textbf{Xe} & & 15.05     & &  4.04  & & -202.3 & & 7.6  & &  0   & & $A^{\prime}$, $A^{\prime\prime}$\\
                                      & &           & &        & & -117.9 & & 7.3  & & 180  & & $A^{\prime}$, $A^{\prime\prime}$\\
                                      & &           & &        & & -224.4 & & 6.45 & & 84   & & $A^{\prime}$       \\
            \hline \hline

            \end{tabular}
        \caption{Properties of the Rg-OH interaction. Values for the potential minima adapted from
        \cite{Gilijamse:Science313:1617,Scharfenberg:PCCP12:10660,Lee:JCP113:5736,Sumiyoshi:PCCP12:8340,Sumiyoshi:MolPhys108:2207}.
        Reduced mass $\mu$ in u, polarizability $\alpha$ in $10^{-24}$cm$^3$, potential energy $E$ in cm$^{-1}$,
        coordinate $R$ in $a_0$, coordinate $\theta$ in degree.}
        \label{tab:pes}
\end{table}

\section{Results and discussion}\label{sec:4}

\subsection{Scattering of OH radicals with Ne, Kr, and Xe atoms}\label{sec:4:1}
In this section, we first describe our new results on the scattering
of OH radicals with Ne, Kr, and Xe atoms. A detailed comparison of
the scattering behavior for the various systems is given in section
\ref{sec:comparison}.

\begin{figure*}[p]
    \centering
    \includegraphics{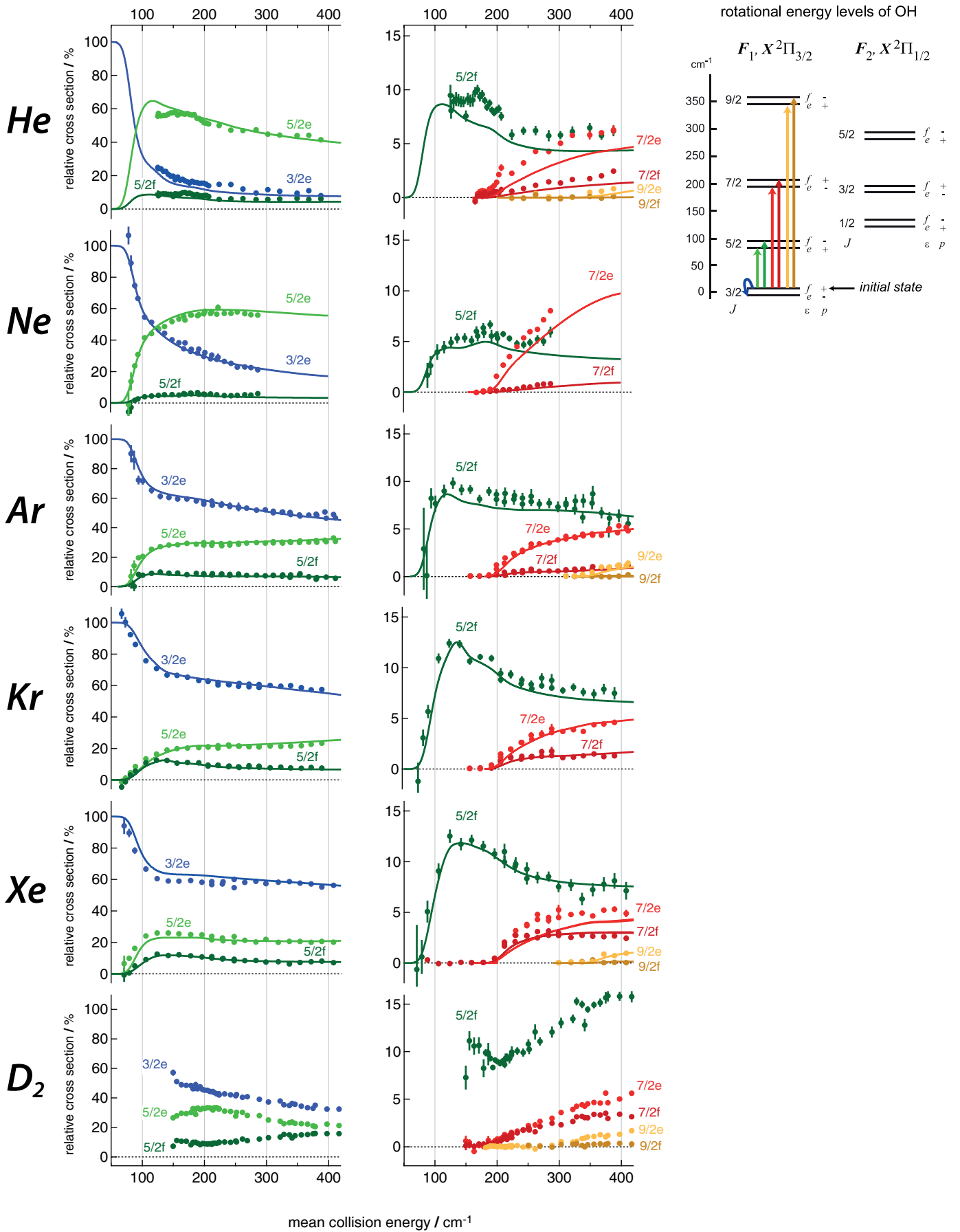}
    \caption{Relative state-to-state inelastic scattering cross sections for spin-orbit conserving ($F_1 \rightarrow F_1$) collisions of
    OH ($X\,^2\Pi_{3/2}, J=3/2, f$) radicals with He, Ne, Ar, Kr, and Xe atoms and D$_2$ molecules as a function of
    the collision energy. The theoretically calculated cross sections are included as solid curves. In the energy-level scheme, the splitting
    between both parity components of each rotational level is largely exaggerated for reasons of clarity. }
    \label{fig:F1_channels}
\end{figure*}

\begin{figure*}[p]
    \centering
  \includegraphics{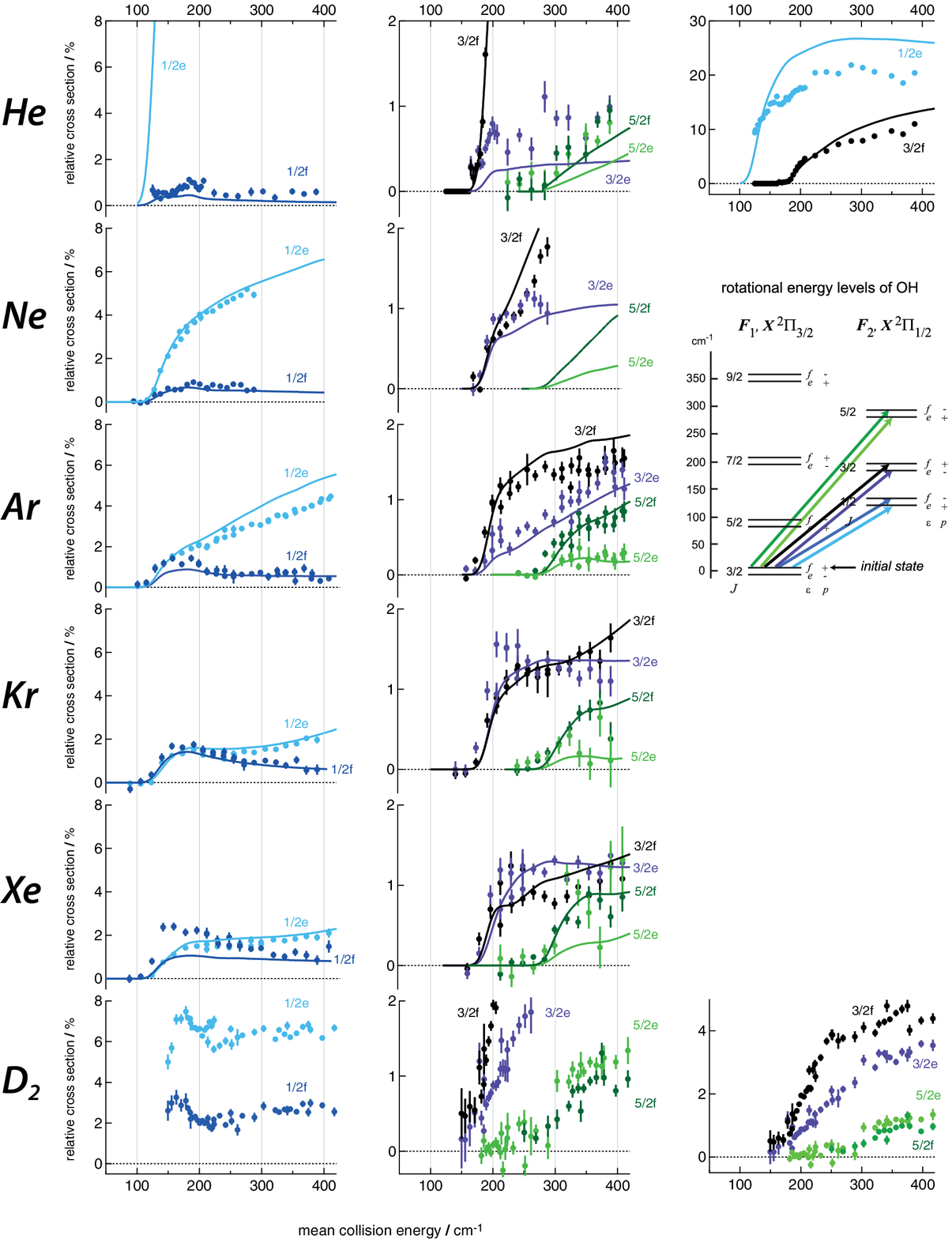}
    \caption{Relative state-to-state inelastic scattering cross sections for spin-orbit changing ($F_1 \rightarrow F_2$) collisions of
    OH ($X\,^2\Pi_{3/2}, J=3/2, f$) radicals with He, Ne, Ar, Kr, and Xe atoms and D$_2$ molecules as a function of
    the collision energy. The theoretically calculated cross sections are included as solid curves.}
    \label{fig:F2_channels}
\end{figure*}

The measured relative state-to-state cross sections for the
scattering of OH radicals with Ne, Kr, and Xe are shown in Figures
\ref{fig:F1_channels} and \ref{fig:F2_channels}. The theoretically
computed cross sections, convoluted with the experimental energy
resolution, are included as solid curves. For completeness, the
experimental and theoretical cross sections from our previous work
for the OH-He, OH-D$_2$, and OH-Ar systems
\cite{Kirste:PRA82:042717,Scharfenberg:PCCP12:10660} are also shown
in these figures. The fine structure conserving (transitions within
the $F_1$ spin-orbit manifold) and fine structure changing
collisions (transitions from the $F_1$ into the $F_2$ manifold) are
summarized in Figure \ref{fig:F1_channels} and Figure
\ref{fig:F2_channels}, respectively. In both figures, the state
resolved scattering channels are labeled following the color codes
as indicated in the rotational energy level diagrams.

For the scattering of OH($F_1(3/2f)$) with Ne, the $\Lambda$-doublet
changing $F_1(3/2f)\rightarrow F_1(3/2e)$ channel dominates at low
collision energies; at energies above $\sim$ 150 cm$^{-1}$ the
scattering is dominated by rotational excitation to the $F_1(5/2e)$
state. For spin-orbit manifold conserving transitions, there is a
strong propensity for final states of $e$ parity. For spin-orbit
manifold changing collisions, a strong $\Lambda$-doublet propensity
is only observed for excitation into the $F_2(1/2)$ states. Note the
little bump that is measured in the cross section for scattering
into the $F_1(5/2f)$ channel at a collision energy just below
200~cm$^{-1}$. This bump is also measured for the OH-He system, and
possibly results from scattering resonances associated with the
opening of the $F_1(7/2)$ channels at a collision energy of
200~cm$^{-1}$. Pronounced scattering resonances are indeed predicted
from the theoretical calculations for OH-He and OH-Ne at these
collision energies, although their specific structure is smeared out
in the experiment due to the collision energy spread.

The scattering behavior of OH with Kr is observed to be very similar
to the scattering of OH with Xe atoms, and dominated by the
$\Lambda$-doublet changing $F_1(3/2f)\rightarrow F_1(3/2e)$ channel
at all probed collision energies. For rotational excitation, the
cross sections generally rise sharply from the energetic threshold,
reach a maximum, and become rather insensitive to a variation of the
collision energy at higher energies. For spin-orbit manifold
conserving collisions, there is a small propensity for excitation
into final states of $e$ parity. For spin-orbit changing collisions,
no clear preference for excitation into one of the $\Lambda$-doublet
components of a final rotational state is observed.

For all three scattering systems, the measured cross sections are
compared to the cross sections determined by quantum close-coupled
calculations based on high quality \emph{ab initio} PES's. For the
scattering of OH with Ne and Kr atoms, excellent agreement is found
between the experimentally determined and theoretically computed
scattering cross sections. The cross sections for all scattering
channels, both for spin-orbit conserving and spin-orbit changing
collisions, and for all collision energies are perfectly reproduced
by the calculations.

For the scattering of OH radicals with Xe atoms, the measured cross
sections confirm the cross sections that were determined in our
previous work on this system \cite{Gilijamse:Science313:1617}. It is
noted that both experiments were performed in different apparatuses
and with different levels of sensitivity. For the present
experiment, in which state-to-state cross sections for a larger
number of final states are measured, excellent agreement is obtained
between experimental and theoretical cross sections. The relative
scattering cross sections, as well as the threshold behavior of
individual channels, are reproduced well. Also the increase in the
relative scattering cross section for the $F_1(3/2f) \rightarrow
F_1(5/2e)$ channel at collision energies just above the energetic
threshold is perfectly reproduced. The only pronounced difference
between experiment and theory is observed for the $F_2(1/2f)$
channel at energies just above threshold. This discrepancy could
possibly be explained by small imperfections in the difference
potential. We have observed in the calculations that the cross
sections for scattering into the $F_2(1/2)$ channels are
particularly sensitive to small variations of $V_{\rm{diff}}$. At
low collision energies, there appears to be a small shift of a few
cm$^{-1}$ between the experimental and the theoretical values for
the collision energy. The origin of this shift is not known, but
could well be the result of an uncertainty in our Xe beam velocity
measurements.

\subsection{Comparison between the various collision
partners}\label{sec:comparison}

Interesting trends are observed when the general scattering behavior
for the various OH-rare gas atom systems are qualitatively compared
with each other. For the series OH-He, Ne, Ar, Kr, Xe it is observed
that the role of the $F_1(3/2f) \rightarrow F_1(3/2e)$ channel
gradually increases. At the same time, propensities for preferred
scattering into the $e$ parity state of the other rotational states
of the $F_1$ manifold tend to get weaker. Finally, the contribution
of the spin-orbit changing $F_1(3/2f) \rightarrow F_2(1/2e)$ channel
to the scattering is gradually reduced from $\sim$ 20 \% for OH-He
to $\sim$ 2~\% for OH-Xe. We note that these qualitative changes are
strongest when the collision partner He is replaced by Ne, and when
Ne is replaced by Ar. When Ar is replaced by Kr, smaller changes are
observed, while hardly any changes occur in going from Kr to Xe. The
scattering of OH radicals with D$_2$ molecules does not fit entirely
in this trend; the overall scattering behavior for this system
resembles that of the OH-Ar, OH-Kr and OH-Xe systems.

A qualitative understanding of the inelastic scattering of OH
radicals with rare-gas atoms can be obtained from a general analysis
given by Dagdigian \emph{et al.} \cite{Dagdigian:JCP91:839}.
According to this analysis, the relative strength of the various
scattering channels (in particular for low values of $J$) can be
estimated from the rotational energy level structure of the OH
radical and the different expansion coefficients $v_{l,0}(R)$ and
$v_{l,2}(R)$ of the sum and difference potential energy surfaces,
respectively.

A close inspection of the nature of the interaction potential and
the relevant coefficients that determine the state-to-state cross
sections can yield a satisfying understanding of the physical origin
of general scattering features \cite{DegliEsposti:JCP103:2067}. As
outlined in Ref. \cite{Kirste:PRA82:042717}, for instance, the
profound difference in the scattering behavior that is observed for
the inelastic scattering of OH with He atoms or D$_2$ molecules can
be explained by the much larger anisotropy of the OH-D$_2$ PES's
compared to the OH-He PES's. In general, a small $F_1(3/2f)
\rightarrow F_1(3/2e)$ $\Lambda$-doublet changing cross section, a
strong propensity for rotational excitation into the $e$ parity
component of the $F_1(5/2)$ state, and a strong cross section for
scattering into the $F_2(1/2e)$ level indicates that the scattering
is dominated by the symmetric $l=$ even terms, whereas the opposite
scattering behavior is expected for systems in which the asymmetric
$l=$ odd terms play a large role. The former is the case for the
weakly interacting OH-He system that results in a potential energy
surface with small anisotropy; the latter applies to the strongly
interacting OH-D$_2$ system.

These qualitative arguments can also be used to rationalize the
trends that are observed for the scattering of OH with the series of
rare-gas atoms He, Ne, Ar, Kr, and Xe. Indeed, with increasing
polarizability of the collision partner, the scattering is governed
by potential energy surfaces with increasing well depth and larger
anisotropy. This results in more dominant $l=$ odd expansion
coefficients, and hence in a larger $F_1(3/2f) \rightarrow
F_1(3/2e)$ $\Lambda$-doublet cross section, smaller $e$ over $f$
propensities for excitation into the $F_1(5/2)$ state, and reduced
cross sections for the $F_1(3/2f) \rightarrow F_2(1/2e)$ spin-orbit
changing channel.

The trends that are observed are thus consistent with what may be
expected from the nature of the OH-rare gas atom interaction
potential. However, for the series of collision partners also
dynamic effects that are related to the increasing mass of the
collision partner may play a role. It is not \emph{a priori} clear
which features in the scattering behavior are due to the nature of
the PES's, and which features result from the atomic and molecular
motion on these PES's.

\begin{figure*}
    \centering
        \includegraphics{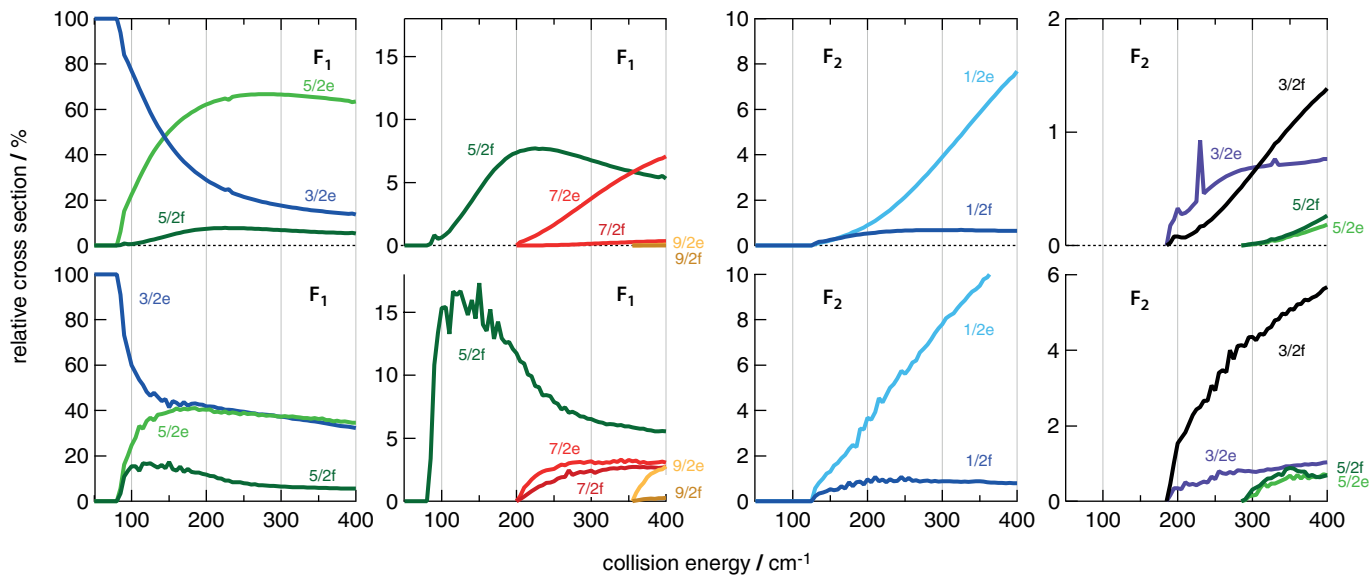}
    \caption{Calculated relative state-to-state inelastic
    scattering cross sections for two hypothetical OH ($X\,^2\Pi_{3/2}, J=3/2, f$)-rare gas atom systems.
    Top row: calculations are based on the OH-He PES's from Ref. \cite{Lee:JCP113:5736}, but the
    scattering calculations are performed with the He atom mass replaced by the Xe atom mass.
    Bottom row: calculations are based on the OH-Xe PES's from Ref. \cite{Gilijamse:Science313:1617}, but the
    scattering calculations are performed with the Xe atom mass replaced by the He atom mass. }
    \label{fig:mass-effects}
\end{figure*}

To study the influence of both parameters on the scattering, we have
performed calculations for two hypothetical OH-rare gas atom
systems. In the first system, we have used the OH-He PES's, but
performed the scattering calculations with the He atom mass replaced
by the Xe atom mass. In the second system, we have used the OH-Xe
PES's, but performed the scattering calculations with the Xe atom
mass replaced by the He atom mass. These model systems thus yield
information on the scattering of OH radicals with a weakly
interacting but heavy collision partner, and with a strongly
interacting but light collision partner. The resulting inelastic
scattering cross sections for both systems are shown in Figure
\ref{fig:mass-effects}.

For the OH-He interaction potential, significant changes in the
cross sections are observed upon replacement of the He atom mass
with the Xe atom mass. In particular, the relative cross section for
scattering into the $F_1(3/2e)$ state increases, while the
contributions of the $F_1(5/2e)$ and $F_2(1/2e)$ channels to the
scattering decrease. Note that also the threshold behavior of
various channels changes significantly; the cross sections rise less
sharply at collision energies above threshold. The resulting
state-to-state relative inelastic cross sections closely resemble
those for the OH-Ne system, i.e., of a weakly interacting system
with a larger reduced mass compared to OH-He.

For the OH-Xe interaction potential, opposite changes in the
scattering cross sections are observed upon replacement of the Xe
mass with the He atom mass. The $F_1(3/2f) \rightarrow F_1(3/2e)$
channel becomes less dominant, and the $F_1(5/2e)$ and $F_2(1/2e)$
channels gain importance. The resulting relative state-to-state
cross sections resemble those that are measured for the OH-D$_2$
system, i.e., of a strongly interacting system with a smaller
reduced mass compared to OH-Xe.

These model calculations indicate that the nature of the potential
energy surface and the reduced mass of the system can both have a
profound and qualitatively similar influence on the scattering cross
sections. This suggests that the interesting trends that are
observed for the scattering of OH with He, Ne, Ar, Kr, and Xe are in
part due to the increasing interaction strength of the OH radical
with the collision partner, and in part due to the increasing mass
of the rare-gas atom. It is not straightforward to disentangle the
influence of both effects using \emph{ab initio} potential energy
surfaces and close-coupled scattering calculations as employed here.
The individual influence on the scattering of properties such as
well depth, anisotropy and reduced mass can be studied best using
models for the potential and the scattering dynamics that allow for
an independent variation of the relevant parameters
\cite{Gijsbertsen:JACS128:8777,Lemeshko:JCP129:024301}.

\section{Conclusions}

We have presented new measurements on the state-to-state rotational
inelastic scattering of Stark-decelerated OH ($X\,^2\Pi_{3/2},
J=3/2, f$) radicals with Ne, Kr, and Xe atoms. For each collision
system, a total of 13~inelastic scattering channels is studied at
collision energies in the $70 - 400$ cm$^{-1}$ range. The collision
energy dependence of the relative inelastic scattering cross
sections, the threshold behavior of inelastic channels, and the
energy dependence of the state-resolved propensities are accurately
determined. Excellent agreement is found with the inelastic
scattering cross sections determined from quantum close-coupled
scattering calculations based on \emph{ab initio} potential energy
surfaces.

These measurements complement our recent studies on the scattering
of the OH radicals with He atoms, Ar atoms, and D$_2$ molecules, and
confirm the measured cross sections of our original work on the
scattering of OH radicals with Xe atoms. Together, these studies
represent the most complete combined experimental and theoretical
study of the inelastic scattering of an open shell radical in a
$^2\Pi$ electronic state with rare-gas atoms. The excellent
agreement that is obtained with the cross sections that are derived
from \emph{ab initio} potential energy surfaces for all OH-rare gas
atom systems clearly indicates that the scattering of these systems
is well understood.

Significant differences are found in the scattering behavior of OH
radicals with the various collision partners, and interesting trends
are observed in the relative inelastic scattering cross sections for
the series OH-He, Ne, Ar, Kr, and Xe. Replacement of the He atom by
heavier rare-gas atoms results in a more dominant $F_1(3/2f)
\rightarrow F_1(3/2e)$ $\Lambda$-doublet changing cross section,
smaller propensities for preferred excitation into one of the
$\Lambda$-doublet components of excited rotational levels, and
reduced cross sections for the $F_1(3/2f) \rightarrow F_2(1/2e)$
transition. These trends result in part from the increasing strength
of the OH-rare gas atom interaction, and in part from the increasing
mass of the rare-gas atom.

The resonances in the scattering of OH with He and Ne atoms at
energies close to threshold -- hinted at in the present measurements
-- will be subject of future experiments with increased collision
energy resolution \cite{Scharfenberg:tobepubl}.

\section{Acknowledgements}
This work is supported by the ESF EuroQUAM programme, and is part of
the CoPoMol (Collisions of Cold Polar Molecules) project. The expert
technical assistance of Georg Hammer, Henrik Haak, Uwe Hoppe and the
FHI mechanical and electronics workshops are gratefully
acknowledged. We thank Anne McCoy and Yoshihiro Sumiyoshi for
providing us the required information to evaluate the fit to the
OH-Ne and OH-Kr potentials, and thank Liesbeth Janssen and Jacek
K{\l}os for help and discussions. AvdA thanks the Alexander von
Humboldt Foundation for a Research Award.


\end{document}